\documentclass{article}
\usepackage{amsmath,amssymb,bbm}
\usepackage{mathrsfs}
\usepackage{cite}

\begin{document}
\title{Negative Energy States in Quantum Theory}
\author{Anastasios Y. Papaioannou \\ \tt{tasosp@gmail.com}}

\maketitle
\begin{abstract}
We analyze the Lagrangian density and canonical stress-energy tensor for the Dirac equation, where the Dirac bispinor has been recast as a multivector set of fields. For the massless Dirac field, the sign of the energy density is determined by the relative phase of uncoupled even- and odd-grade field components. These components become coupled in the massive Dirac equation, and the sign of the energy is determined by their spatial parity. The corresponding stress-energy tensors for the second-order equations also admit negative energy states, with the sign of the energy density again dependent on field parity.

We apply the same multivector approach to electromagnetism, constructing new Lagrangian and energy densities in which the vector potential and the electromagnetic field are treated as independent field degrees of freedom.
\end{abstract}
\section{The massless Dirac field}
In previous work \cite{papaioannou2017}, \cite{papaioannou2018}, we derived multivector-valued equivalents of the massless and massive Dirac field equations, showing that the spinor degrees of freedom can always be separated from the physical scalar, vector, bivector, pseudovector, and pseudoscalar field degrees of freedom. The massless Dirac equation
\begin{equation*}
j \gamma_\mu \partial^\mu \psi = 0
\end{equation*}
takes the multivector form
\begin{equation*}
j \nabla M = 0,
\end{equation*}
where
\begin{equation*}
\nabla = e_\mu \partial^\mu,
\end{equation*}
and the complex multivector $M$ can be split into real even-grade and imaginary odd-grade terms,
\begin{equation*}
M = M_e + j M_o.
\end{equation*}
$M_e$ contains the even-grade (scalar, bivector, and pseudoscalar) terms,
\begin{equation*}
M_e = f \, 1 + F + g I,
\end{equation*}
while $M_o$ contains the odd-grade (vector and pseudovector) terms,
\begin{equation*}
M_o = A + I p.
\end{equation*}
We write the fields in component form as
\begin{align*}
F & = E^k e_k e_0 + B^k I e_k e_0 \\
A & = \varphi e_0 + A^k e_k \\
I p & = \chi I e_0 + p^k I e_k,
\end{align*}
with
\begin{equation*}
I \equiv e_0 e_1 e_2 e_3.
\end{equation*}
Unlike in other references in the literature (e.g., \cite{papaioannou2017}, \cite{hestenes-local-observables}, \cite{doran-lasenby}), we do not assign a geometric interpretation to the imaginary unit scalar $j$, as that would require the introduction of a preferred direction in spacetime.

The Lagrangian density
\begin{equation*}
\mathcal{L} = \text{Re} \{ j \overline{\psi} \gamma^\mu \partial_\mu \psi \}
\end{equation*}
takes the form
\begin{equation*}
\mathcal{L} = \langle \widetilde{M}_o \nabla M_e - \widetilde{M}_e \nabla M_o \rangle,
\end{equation*}
where $\widetilde{M}$ denotes the multivector reversal of each term in $M$, e.g.,
\begin{equation*}
(e_{\alpha_1} e_{\alpha_2} \cdots e_{\alpha_k})^\sim = e_{\alpha_k} \cdots e_{\alpha_2} e_{\alpha_1},
\end{equation*}
and $\langle M \rangle$ denotes the term in $M$ proportional to the scalar element $1$. In component form, the Lagrangian density is
\begin{align*}
\mathcal{L} = 
& + \varphi \, (\partial_t f + \vec{\nabla} \cdot \vec{E}) + \chi \, (\partial_t g + \vec{\nabla} \cdot \vec{B}) \\
& + \vec{A} \cdot (\partial_t \vec{E} - \vec{\nabla} \times \vec{B} + \vec{\nabla} f) + \vec{p} \cdot (\partial_t \vec{B} + \vec{\nabla} \times \vec{E} + \vec{\nabla} g) \\
& -f \, (\partial_t \varphi + \vec{\nabla} \cdot \vec{A}) - g \, (\partial_t \chi + \vec{\nabla} \cdot \vec{p}) \\
& - \vec{E} \cdot (\partial_t \vec{A} + \vec{\nabla} \varphi - \vec{\nabla} \times \vec{p}) - \vec{B} \cdot (\partial_t \vec{p} + \vec{\nabla} \chi + \vec{\nabla} \times \vec{A}).
\end{align*}
The corresponding Euler-Lagrange equations
\begin{align*}
\nabla M_e & = 0 \\
\nabla M_o & = 0
\end{align*}
become
\begin{align*}
\partial_t f + \vec{\nabla} \cdot \vec{E} & = 0 \\
\partial_t g + \vec{\nabla} \cdot \vec{B} & = 0 \\
\partial_t \vec{E} - \vec{\nabla} \times \vec{B} + \vec{\nabla} f & = 0 \\
\partial_t \vec{B} + \vec{\nabla} \times \vec{E} + \vec{\nabla} g & = 0 \\
\partial_t \varphi + \vec{\nabla} \cdot \vec{A} & = 0 \\
\partial_t \chi + \vec{\nabla} \cdot \vec{p} & = 0 \\
\partial_t \vec{A} + \vec{\nabla} \varphi - \vec{\nabla} \times \vec{p} & = 0 \\
\partial_t \vec{p} + \vec{\nabla} \chi + \vec{\nabla} \times \vec{A} & = 0.
\end{align*}
The corresponding canonical stress-energy density is
\begin{equation*}
T^{\mu \nu} = \text{Re} \left\{ j \overline{\psi} \gamma^\mu \partial^\nu \psi \right\} - \mathcal{L} g^{\mu \nu}.
\end{equation*}
For all solutions of the Euler-Lagrange equations, $\mathcal{L}$ is identically zero, and the above expression simplifies to
\begin{equation*}
T^{\mu \nu} = \text{Re} \left\{ j \overline{\psi} \gamma^\mu \partial^\nu \psi \right\},
\end{equation*}
or,
\begin{align*}
T^{\mu \nu} & = \langle - \widetilde{M}_e e^\mu \partial^\nu M_o + \widetilde{M}_o e^\mu \partial^\nu M_e \rangle \\
& = \langle (- \partial^\nu M_o \widetilde{M}_e + \partial^\nu M_e \widetilde{M}_o) e^\mu \rangle.
\end{align*}
Although $M_e$ and $M_o$ are not coupled by the Euler-Lagrange equations, they are coupled in the stress-energy tensor, and for every positive-energy state
\begin{equation*}
M_+ = M_e + j M_o
\end{equation*}
there is a valid negative-energy state
\begin{equation*}
M_- = M_+^*.
\end{equation*}
For example, the plane-wave solution
\begin{align*}
M_e & = f \, 1 + E^3 e_3 e_0 \\
M_o & = \varphi \, e_0 + A^3 e_3,
\end{align*}
with
\begin{align*}
f & = E^3 = a_0 \cos(k z - \omega t) \\
\varphi & = A^3 = a_0 \sin (k z - \omega t),
\end{align*}
has a positive, uniform, constant energy density
\begin{align*}
u & = + \varphi \, \partial_t f - f \, \partial_t \varphi + \chi \, \partial_t g - g \, \partial_t \chi \\
& + \vec{A} \cdot \partial_t \vec{E} - \vec{E} \cdot \partial_t \vec{A} + \vec{p} \cdot \partial_t \vec{B} - \vec{B} \cdot \partial_t \vec{p} \\
& = \varphi \, \partial_t f - f \, \partial_t \phi + A^3 \, \partial_t E^3 - E^3 \, \partial_t A^3 \\
& = + 2 a^2_0 \omega,
\end{align*}
while the solution
\begin{align*}
f & = E^3 = a_0 \cos(k z - \omega t) \\
\varphi & = A^3 = - a_0 \sin (k z - \omega t),
\end{align*}
has an energy density of the same magnitude but opposite sign,
\begin{equation*}
u = - 2 a^2_0 \omega.
\end{equation*}
Compare this to the Lagrangian density for the second-order multivector-valued massless wave equation:
\begin{equation*}
\mathcal{L} = \frac{1}{2} \langle (\nabla M_e) (\nabla M_e)^\sim + (\nabla M_o) (\nabla M_o)^\sim \rangle.
\end{equation*}
Cross-terms in the Lagrangian density between differing field components do not contribute to the equations of motion. We can therefore construct a simplified Lagrangian density which contains only the diagonal terms, yet nevertheless yields the same set of second-order differential equations:
\begin{align*}
\mathcal{L} & = \frac{1}{2} \langle \partial^\alpha M_e \partial_\alpha \widetilde{M}_e + \partial^\alpha M_o \partial_\alpha \widetilde{M}_o \rangle \\
& = \frac{1}{2} (\partial_\alpha f \partial^\alpha f + \partial_\alpha \varphi \, \partial^\alpha \varphi + \partial_\alpha \vec{B} \cdot \partial^\alpha \vec{B} + \partial_\alpha \vec{p} \cdot \partial^\alpha \vec{p}) \\
& - \frac{1}{2} (\partial_\alpha g \, \partial^\alpha g + \partial_\alpha \chi \, \partial^\alpha \chi + \partial_\alpha \vec{E} \cdot \partial^\alpha \vec{E} + \partial_\alpha \vec{A} \cdot \partial^\alpha \vec{A}).
\end{align*}
The positive-parity components
\begin{equation*}
f, \, \varphi, \, \vec{B}, \, \vec{p}
\end{equation*}
all contribute positive values to the energy density, e.g.,
\begin{equation*}
u_f = + \frac{1}{2} (\partial_t f)^2 + \frac{1}{2} (\vec{\nabla} f)^2,
\end{equation*}
while the negative-parity components
\begin{equation*}
g, \, \chi, \, \vec{E}, \, \vec{A}
\end{equation*}
all have negative energy, e.g.,
\begin{equation*}
u_g = - \frac{1}{2} (\partial_t g)^2 - \frac{1}{2} (\vec{\nabla} g)^2.
\end{equation*}
\section{The massive Dirac field}
With the introduction of the mass term, the Dirac equation
\begin{equation*}
j \nabla M = \omega_0 M
\end{equation*}
couples $M_e$ and $M_o$:
\begin{align*}
\nabla M_e & = + \omega_0 M_o \\
\nabla M_o & = - \omega_0 M_e.
\end{align*}
$M$ is a solution of the Klein-Gordon equation, as expected:
\begin{align*}
(j \nabla) (j \nabla) M & = - \Box M \\
& = + \omega^2_0 M.
\end{align*}
Conversely, we can derive a Dirac solution given a Klein-Gordon solution. Suppose we have a real multivector-valued Klein-Gordon field:
\begin{equation*}
\Box M_K = -\omega^2_0 M_K.
\end{equation*}
Then the complex multivector
\begin{align*}
M & \equiv M_K + j M'_K, \\
M'_K & \equiv \frac{1}{\omega_0} \nabla M_K
\end{align*}
satisfies the Dirac equation:
\begin{align*}
\nabla M_K & = + \omega_0 M'_K \\
\nabla M'_K & = - \omega_0 M_K.
\end{align*}
If $M$ obeys the initial condition
\begin{equation*}
M(t=t_0) = A,
\end{equation*}
then the time evolution of $M$ for a particle at rest is
\begin{align*}
\partial_t M & = -j e_0 \omega_0 M \\
\to M(t) & = e^{-j e_0 \omega_0 (t - t_0)} A.
\end{align*}
We recover the familiar bispinor solution by representing $e_\mu$ with the gamma matrices $\gamma_\mu$ and right-multiplying the field by a fixed bispinor, which takes the form
\begin{equation*}
w = \begin{pmatrix} 1 \\ 0 \\ 0 \\ 0  \end{pmatrix}
\end{equation*}
in the rest frame:
\begin{equation*}
\psi(t) = e^{-j e_0 \omega_0 t} A \, w.
\end{equation*}
The spatial parity of the multivector determines the sign of the energy: for a parity eigenstate $A$ with eigenvalue $P$,
\begin{equation*}
e_0 A = (-1)^P A e_0,
\end{equation*}
and
\begin{align*}
\psi(t) & = A \, e^{-j (-1)^P e_0 \omega_0 t} \, w \\
& = A \, w \, e^{-j (-1)^P \omega_0 t}.
\end{align*}
We see that the positive-parity multivectors
\begin{equation*}
A_+ \in \{ 1, e_0, I e_i e_0, I e_i \}
\end{equation*}
correspond to the positive-energy states
\begin{equation*}
\psi_+(t) = A_+ w \, e^{-j \omega_0 t},
\end{equation*}
and the negative-parity multivectors
\begin{equation*}
A_- \in \{ I, e_i, e_i e_0, I e_0 \}
\end{equation*}
correspond to the negative-energy states
\begin{equation*}
\psi_-(t) = A_- w \, e^{+j \omega_0 t}.
\end{equation*}
The Lagrangian density becomes
\begin{align*}
\mathcal{L} & = \mathcal{L}_0 - \omega_0 \overline{\psi} \psi \\
& \to \mathcal{L}_0 - \omega_0 \langle M_e \widetilde{M}_e + M_o \widetilde{M}_o \rangle \\
& = \mathcal{L}_0 - \omega_0 (f^2 + \varphi^2 + \vec{p}^2 + \vec{B}^2) + \omega_0 (g^2 + \chi^2 + \vec{A}^2 + \vec{E}^2),
\end{align*}
where $\mathcal{L}_0$ is the Lagrangian density of the massless field. The signs of the mass terms reflect the field parities, as expected.

\section{A new Lagrangian for electromagnetism}
We can make the close resemblance between the component-form Dirac equations and Maxwell's equations exact, with a slight modification to the massless Lagrangian density:
\begin{align*}
\mathcal{L} = & \, \mathcal{L}_0 + \langle F \widetilde{F} \rangle \\
= & \, \mathcal{L}_0 + (\vec{B}^2 - \vec{E}^2) \\
= & + \varphi \, (\partial_t f + \vec{\nabla} \cdot \vec{E}) + \chi \, (\partial_t g + \vec{\nabla} \cdot \vec{B}) \\
& + \vec{A} \cdot (\partial_t \vec{E} - \vec{\nabla} \times \vec{B} + \vec{\nabla} f) + \vec{p} \cdot (\partial_t \vec{B} + \vec{\nabla} \times \vec{E} + \vec{\nabla} g) \\
& - f \, (\partial_t \varphi + \vec{\nabla} \cdot \vec{A}) - g \, (\partial_t \chi + \vec{\nabla} \cdot \vec{p}) \\
& - \vec{E} \cdot (\partial_t \vec{A} + \vec{\nabla} \varphi - \vec{\nabla} \times \vec{p} + \vec{E}) - \vec{B} \cdot (\partial_t \vec{p} + \vec{\nabla} \chi + \vec{\nabla} \times \vec{A} - \vec{B}).
\end{align*}
The Euler-Lagrange equations,
\begin{align*}
\partial_t f + \vec{\nabla} \cdot \vec{E} & = 0 \\
\partial_t g + \vec{\nabla} \cdot \vec{B} & = 0 \\
\partial_t \vec{E} - \vec{\nabla} \times \vec{B} + \vec{\nabla} f & = 0 \\
\partial_t \vec{B} + \vec{\nabla} \times \vec{E} + \vec{\nabla} g & = 0 \\
\partial_t \varphi + \vec{\nabla} \cdot \vec{A} & = 0 \\
\partial_t \chi + \vec{\nabla} \cdot \vec{p} & = 0 \\
\partial_t \vec{A} + \vec{\nabla} \varphi - \vec{\nabla} \times \vec{p} + \vec{E} & = 0 \\
\partial_t \vec{p} + \vec{\nabla} \chi + \vec{\nabla} \times \vec{A} - \vec{B} & = 0,
\end{align*}
become Maxwell's equations when the scalar, pseudovector, and pseudoscalar fields vanish,
\begin{equation*}
f = g = \chi = \vec{p} = 0.
\end{equation*}
The potentials obey the Lorenz gauge condition,
\begin{equation*}
\partial_t \varphi + \vec{\nabla} \cdot \vec{A} = 0,
\end{equation*}
and the equations
\begin{align*}
\vec{E} & = - \vec{\nabla} \varphi - \partial_t \vec{A} \\
\vec{B} & = \vec{\nabla} \times \vec{A}
\end{align*}
are dynamic equations of motion and not definitions of $\vec{E}$ and $\vec{B}$.

Associated with this new Dirac-type Lagrangian density is a new on-shell conserved energy density,
\begin{align*}
u & = \vec{A} \cdot \partial_t \vec{E} - \vec{E} \cdot \partial_t \vec{A} \\
& = \vec{A} \cdot (\vec{\nabla} \times \vec{B}) + \vec{E} \cdot (\vec{\nabla} \varphi + \vec{E}),
\end{align*}
distinct from the conventional quantity
\begin{equation*}
u_c = \frac{1}{2} ( \vec{E}^2 + \vec{B}^2 ).
\end{equation*}
For example, a traveling mode solution
\begin{align*}
E^1 & = a_0 \cos(k z - \omega t) \\
B^2 & = a_0 \cos(k z - \omega t) \\
A^1 & = -\frac{a_0}{\omega} \sin(k z - \omega t),
\end{align*}
has a Dirac-type energy density that is uniform and constant:
\begin{equation*}
u = + a^2_0 \omega,
\end{equation*}
while its conventional energy density is neither:
\begin{equation*}
u_c = a^2_0 \cos^2(k z - \omega t),
\end{equation*}
although both are positive-definite.

\bibliography{negative-energy-states-in-quantum-theory}
\bibliographystyle{hunsrt}

\end{document}